\documentclass[twocolumn]{article}
\usepackage[margin=0.6in,columnsep=0.15in]{geometry}
\usepackage{upgreek}
\usepackage{dcolumn}
\usepackage[utf8]{inputenc}

\usepackage[usenames,dvipsnames]{xcolor}
\PassOptionsToPackage{hyphens}{url}
\usepackage[colorlinks,citecolor=Gray,urlcolor=Gray,linkcolor=Blue]{hyperref}

\usepackage{graphicx}
\usepackage{mathtools}
\usepackage{booktabs}
\usepackage{tabularx}
\usepackage{tikz}
\usepackage[multi-part-units=single]{siunitx}
\DeclareSIUnit{\calorie}{cal}
\DeclareSIUnit{\kcal}{\kilo\calorie}
\DeclareSIUnit{\bohr}{\text{\ensuremath{a}}_{0}}
\usepackage{dcolumn}
\usepackage{multirow}
\usepackage[textsize=footnotesize]{todonotes}
\usepackage{float}
\usepackage{makecell}
\usepackage[capitalize]{cleveref}
\usepackage{listings}
\crefname{lstlisting}{Listing}{Listings}
\DeclareUnicodeCharacter{2713}\checkmark%

\usepackage[lcgreekalpha,upint]{stix}
\usepackage[scaled=0.8]{FiraMono}

\newcommand{\eqlabelleft}{(}
\newcommand{\eqlabelright}{)}
\DeclareRobustCommand{\pcref}[1]{%
\begingroup%
\renewcommand{\eqlabelleft}{}%
\renewcommand{\eqlabelright}{}%
\cref{#1}%
\endgroup%
}
\creflabelformat{equation}{\eqlabelleft#2#1#3\eqlabelright}

\usepackage[tiny]{titlesec}

\usepackage[inline]{enumitem}
\setlist{nosep}

\usepackage[normal]{caption}
\DeclareCaptionLabelSeparator{pipe}{ $|$ }
\usepackage{authblk}

\usepackage{ifpdf}
\ifpdf%
\usepackage[style=authoryear-comp]{biblatex}
\else
\usepackage[style=authoryear-comp,backend=bibtex]{biblatex}
\fi
\let\citep\autocite%
\let\citet\textcite%
\AtBeginBibliography{\footnotesize}

\title{\Large libMBD: A general-purpose package for scalable quantum
many-body dispersion calculations}

\author[1,*,\(\dagger\)]{Jan Hermann}
\author[2]{Martin Stöhr}
\author[2]{Szabolcs Góger}
\author[3]{Shayantan Chaudhuri}
\author[4]{Bálint Aradi}
\author[3,5]{Reinhard J. Maurer}
\author[2,*]{Alexandre Tkatchenko}
\affil[1]{FU Berlin, Department of Mathematics and Computer Science,
14195 Berlin, Germany}
\affil[2]{University of Luxembourg, Department of Physics and Materials
Science, L-1511 Luxembourg City, Luxembourg}
\affil[3]{University of Warwick, Department of Chemistry, Coventry, CV4
7AL, United Kingdom}
\affil[4]{University of Bremen, Bremen Center for Computational
Materials Science, 28359 Bremen, Germany}
\affil[5]{University of Warwick, Department of Physics, Coventry, CV4
7AL, United Kingdom}

\date{}

\makeatletter
\def\blfootnote{\xdef\@thefnmark{}\@footnotetext}
\makeatother

\begin{document}

\twocolumn[{%
  \maketitle
  \vspace{-3em}
  \begin{center}
  \begin{minipage}{0.85\linewidth}
    \small
    \paragraph{Abstract}
    Many-body dispersion (MBD) is a powerful framework to treat van der
    Waals (vdW) dispersion interactions in density-functional theory and
    related atomistic modeling methods. Several independent
    implementations of MBD with varying degree of functionality exist
    across a number of electronic structure codes, which both limits the
    current users of those codes and complicates dissemination of new
    variants of MBD. Here, we develop and document libMBD, a library
    implementation of MBD that is functionally complete, efficient, easy
    to integrate with any electronic structure code, and already
    integrated in FHI-aims, DFTB+, VASP, Q-Chem, CASTEP, and Quantum
    ESPRESSO. libMBD is written in modern Fortran with bindings to C and
    Python, uses MPI/ScaLAPACK for parallelization, and implements MBD
    for both finite and periodic systems, with analytical gradients with
    respect to all input parameters. The computational cost has
    asymptotic cubic scaling with system size, and evaluation of
    gradients only changes the prefactor of the scaling law, with libMBD
    exhibiting strong scaling up to 256 processor cores. Other MBD
    properties beyond energy and gradients can be calculated with
    libMBD, such as the charge-density polarization, first-order Coulomb
    correction, the dielectric function, or the order-by-order expansion
    of the energy in the dipole interaction. Calculations on
    supramolecular complexes with MBD-corrected electronic structure
    methods and a meta-review of previous applications of MBD
    demonstrate the broad applicability of the libMBD package to treat
    vdW interactions.
  \end{minipage}
  \end{center}
  \vspace{1em}
}]

\blfootnote{\hspace{-0.8em}$^*$Emails: \href{mailto:science@jan.hermann.name}{science@jan.hermann.name}, \href{mailto:alexandre.tkatchenko@uni.lu}{alexandre.tkatchenko@uni.lu}

$^\dagger$Current address: \textit{Microsoft Research AI4Science, Karl-Liebknecht-Str.\ 32, 10178 Berlin, Germany}}%


\section{Introduction}

Van der Waals (vdW) dispersion interactions stem from electron correlation induced by the long-range part of the Coulomb force.
Since they are attractive under normal circumstances, their importance grows with the characteristic length scale in an atomistic system, making them an often decisive force in large molecular complexes, molecular crystals, nanostructured materials, surface phenomena, and soft matter in general.
At the same time, the workhorse electronic structure method for simulating molecules and materials---Kohn--Sham density-functional theory (KS-DFT) in semilocal and hybrid approximations---neglects vdW interactions by construction.
As a result, a large number of approaches to mitigate this issue have been developed, and some of them are now used routinely when performing DFT calculations \citep{HermannCR17,GrimmeCR16}.

One such popular approach is the many-body dispersion (MBD) method, which is based on a model Hamiltonian of charged harmonic oscillators that capture the long-range electrodynamic response of an atomic system and yield its vdW energy \citep{TkatchenkoPRL12}.
Its essential offering is that the coarse-graining of the full electronic structure to oscillators makes it efficient, while full many-body treatment of the electronic fluctuations provides consistent description of beyond-pairwise vdW interactions \citep{AmbrosettiS16}.
Both these characteristics make the MBD framework especially apt for large-scale modeling of molecular systems, for which the demand is bound to only increase in the future \citep{AkimovCR15}.
Given the existence of MBD integrations with even more efficient substitutes of DFT based on density-functional tight binding (DFTB) \citep{StohrJCP16} and machine learning \citep{BereauJCP18,WestermayrDD22,PoierJPCL22,UnkeCR21}, as well as the fact that MBD is an approach that is still evolving and being improved and extended (see below), the situation calls for an efficient, flexible, and reusable implementation of MBD that can be easily deployed in any given electronic structure code.
Here, we develop and document libMBD, a library package that answers this call.

\citet{Libmbd} is written in modern Fortran with additional bindings available for C and Python.
It uses \citet{MPI}/\citet{ScaLAPACK} parallelization, and implements MBD for both finite and periodic systems, with analytical gradients of the energy with respect to all input parameters, which enables both structure optimization as well as self-consistent vdW energy with respect to the electron density \citep{FerriPRL15}.
libMBD is designed as a framework for any method based on the MBD Hamiltonian, and has a built-in support for the original MBD formulation \citep{TkatchenkoPRL12}, the range-separated self-consistently screened variant (MBD@rsSCS) \citep{AmbrosettiJCP14}, the metal-surface parameterizations (MBD$^\text{surf}$) \citep{RuizPRL12,RuizPRB16}, as well as the recent hybrid of MBD with nonlocal functionals (MBD-NL) \citep{HermannPRL20}.
For convenience, an efficient implementation of the predecessor of MBD, the pairwise TS model \citep{TkatchenkoPRL09}, is included as well.
Finally, libMBD offers not only the vdW energy and its gradients, but can calculate also other quantities derived from the MBD Hamiltonian, namely the eigenstate charge fluctuations and induced charge-density polarization \citep{HermannNC17}, first-order Coulomb corrections \citep{StohrNC21}, the dielectric function \citep{CuiJPCL20}, and the order-by-order expansion of the energy in the dipole interaction \citep{TkatchenkoJCP13}.

The popularity of MBD as a method is revealed through its independent implementations in numerous electronic structure codes, namely \citet{FHI-aims} \citep{BlumCPC09}, \citet{VASP} \citep{KressePRB96}, \citet{Q-Chem} \citep{EpifanovskyJCP21}, \citet{Quantum-ESPRESSO} \citep{GiannozziJCP20}, \citet{CASTEP} \citep{ClarkZFK05}, and \citet{DFT-D4} \citep{CaldeweyherJCP19}.
The original standalone reference implementation of MBD is available for download via Internet Archive \citep{MBD}.
While some of these implementations introduced methodological innovations---such as the one in VASP, which introduced a reciprocal-space formulation \citep{BuckoJPCM16}, and the one in Quantum ESPRESSO, which introduced analytical gradients \citep{Blood-ForsytheCS16}---all of them lack both in computational efficiency and in capability compared to libMBD.
The single exception is the linear-scaling stochastic evaluation of MBD \citep{PoierJCTC22}, which is more efficient for very large systems, but for the moment lacks analytical gradients.
Furthermore, none of these MBD implementations have been designed as a library and their reuse across different electronic structure codes is problematic in theory, and nonexistent in practice.
In contrast, libMBD has been already integrated into several such codes, including FHI-aims, VASP, Q-Chem, Quantum ESPRESSO, DFTB+ \citep{HourahineJCP20}, and \citet{ASE} \citep{LarsenJPCM17}.

The flexibility of MBD as a general approach can be seen in the ever-growing list of methods and models that are based on it, and for this reason libMBD has been designed from the start as a framework that is easy to extend and modify.
Besides the methods listed above that are already incorporated in libMBD, there are the fractionally ionic variant (MBD/FI) \citep{GouldJCTC16a} and its successor ``universal'' MBD (uMBD) \citep{KimJACS20}, the Wannier-function parameterization (vdW-WF2) \citep{SilvestrelliJCP13}, beyond-dipole extension \citep{MassaES21}, anisotropic extension \citep{BandyopadhyayPCCP22}, extension to excited states \citep{AmbrosettiNC22}, and the MBD variant of DFT-D4 \citep{CaldeweyherJCP19}.
If implemented in or with libMBD, these methods and models or any future ones would become instantly available across all the electronic structure codes to which libMBD has already been or will be integrated, and could  take advantage of parallelization, analytical gradients, and both finite and periodic formulations.
Next to computational efficiency, this is the biggest offering of libMBD.

The rest of the paper is structured as follows:
\Cref{sec:theory} reviews all the MBD methods implemented in libMBD in terms of equations, as well as the physics of integrating MBD with DFT and its substitutes.
\Cref{sec:implementation} documents the structure of the code and the public interface served to the electronic structure codes, its scaling properties, numerical convergence, functionality, and existing interfaces.
\Cref{sec:applications} describes a number of example MBD calculations and a meta-review of important past applications of MBD.
The paper is concluded in \cref{sec:conclusions}.

\section{Theory}
\label{sec:theory}

This section reviews all mathematical equations that define most of the functionality of libMBD.
It is structured such that equations in \crefrange{sec:mbd}{sec:mbd-methods} are implemented in libMBD, while equations in \cref{sec:integration} must be implemented in the codes with which libMBD is integrated, since their implementation is inherently code-specific.

\subsection{Many-body dispersion}
\label{sec:mbd}

Any MBD method maps an atomic system to a model Hamiltonian of charged harmonic (Drude) oscillators located at $\mathbf R_i$, with static polarizabilities $\alpha_{0,i}$ and frequencies $\omega_i$.
In most variants, the oscillators represent atoms, but both finer fragments, such as Wannier functions \citep{SilvestrelliJCP13}, and coarser fragments, such as whole molecules \citep{JonesPRB13}, are also possible.
As a result of the coarse-graining, only the long-range part of the interaction between oscillators is expected to represent faithfully the original electronic system, justifying a dipole approximation.
The MBD Hamiltonian for a finite system of $N$ oscillators then reads
\begin{equation}
\begin{multlined}
  H_\text{MBD}(\{\mathbf R_i,\alpha_{0,i},\omega_i,\boldsymbol\kappa_i\})
  :=\sum_i-\frac12\boldsymbol\nabla_{\xi_i}^2+\sum_i\frac12\omega_i^2\xi_i^2 \\
  +\frac12\sum_{i\neq j}\omega_i\omega_j\sqrt{\alpha_{0,i}\alpha_{0,j}}
  \boldsymbol\xi_i\cdot\mathbf T^\mathrm{lr}(\boldsymbol\kappa_i,\boldsymbol\kappa_j,\mathbf R_j-\mathbf R_i)\boldsymbol{\xi}_j
\end{multlined}\label{eq:mbd-hamiltonian}
\end{equation}
Here, $\boldsymbol\xi_i=\sqrt{m_i}(\mathbf r_i-\mathbf R_i)$ are displacements of the oscillating charges weighted by the oscillator masses $m_i$, and $\mathbf T^\mathrm{lr}$ is a damped dipole interaction tensor, parametrized through some single-oscillator properties $\boldsymbol\kappa_i$.
The mass is in principle the third parameter (after static polarizability and frequency) required to fully specify a charged quantum oscillator, but it has no effect on the energy under the dipole approximation.
Other sets of three independent oscillator parameters can be used as well, for example the $C_6$ dispersion coefficient or charge $q$, but these are always connected through simple relations, such as
\begin{equation}
4C_6=3\omega\alpha_0^2,\qquad q^2=m\alpha_0\omega^2
\label{eq:osc-param}
\end{equation}
The damped dipole tensor must reduce to the ``bare'' dipole tensor $\mathbf T$ for large distances,
\begin{gather}
\lim_{r\rightarrow\infty}\mathbf T^\mathrm{lr}(\boldsymbol\kappa,\boldsymbol\kappa',\mathbf r)=\mathbf T(\mathbf r), \\
T_{ab}(\mathbf r)=\frac{\partial^2}{\partial r_a\partial r_b}\frac1{\lvert\mathbf r\rvert}=
\frac{-3r_a r_b+\delta_{ab}r^2}{r^5} \label{eq:dipole}
\end{gather}
Within this general MBD framework, a particular MBD model is then specified through the level of coarse-graining (typically atoms), the parametrization of the oscillators, and the choice of $\mathbf T^\text{lr}$, which necessarily depends on the effective range (degree of non-locality) of the electronic structure method to which MBD is coupled.

A major consequence of the dipole approximation is that the MBD Hamiltonian can be exactly diagonalized, after which it describes a system of uncoupled collective oscillations $\tilde\xi_k=\sum_{ia}C_{ia,k}\xi_{ia}$.
Here, we use a shorthand notation in which the particle and Cartesian indexes are joined, for example ${(\boldsymbol\xi_i)}_a=\xi_{ia}={(\boldsymbol\xi)}_{ia}$ and ${(\mathbf C_{ij})}_{ab}=C_{ia,jb}={(\mathbf C)}_{ia,jb}$.
The MBD energy is then obtained as the change in the zero-point energy of the oscillations (fluctuations) induced by the dipole interaction,
\begin{gather}
\mathbf Q_{ij}=\delta_{ij}\omega_i^2\mathbf I+
\omega_i\omega_j\sqrt{\alpha_{0,i}\alpha_{0,j}}\mathbf T_{ij}^\text{lr} \label{eq:q-matrix} \\
\mathbf Q\equiv
\mathbf C\boldsymbol\Lambda\mathbf C^\dagger,\qquad
\boldsymbol\Lambda\equiv\operatorname{diag}(\tilde\omega_1^2,\ldots,\tilde\omega_{3N}^2) \label{eq:diagonalization} \\
E_\text{MBD}=\sum_{k=1}^{3N}\frac{\tilde\omega_k}2 - 3\sum_i\frac{\omega_i}2 \label{eq:mbd-energy}
\end{gather}
with $\mathbf T_{ij}^\text{lr}\equiv\mathbf T^\text{lr}(\boldsymbol\kappa_i,\boldsymbol\kappa_j,\mathbf R_j-\mathbf R_i)$.
Note that the diagonalization leads to complex-valued energy if some of the eigenvalues of $\mathbf Q$ are negative.
This can result from a failure of the dipole approximation, and in typical scenarios signifies that the particular parametrization of the MBD Hamiltonian does not properly represent the local polarization response of a given molecule or material.

\subsection{Periodic boundary conditions}
\label{sec:pbc}

In a periodic system, the collective oscillations have an associated wave vector $\mathbf q$ from the first Brillouin zone (FBZ) as a good quantum number, and the energy per periodic unit cell can be calculated as an integral over the FBZ, evaluated in practice as a sum over a $\mathbf q$-point mesh,
\begin{equation}
E_\text{MBD}=\int_\text{FBZ}\mathrm d\mathbf q\,E_\text{MBD}(\mathbf q)
\label{eq:brillouin-int}
\end{equation}
$E_\text{MBD}(\mathbf q)$ is obtained from \crefrange{eq:q-matrix}{eq:mbd-energy} by making $\tilde{\boldsymbol\omega}$, $\mathbf T^\text{lr}$, and $\mathbf C$ formally dependent on $\mathbf q$ and considering
\begin{equation}
\mathbf T_{ij}(\mathbf q)=\sideset{}{'}\sum_{\mathbf n}\mathbf T(\mathbf R_{\mathbf
nij})\mathrm e^{-\mathrm i\mathbf q\cdot\mathbf R_{\mathbf nij}},\quad\mathbf
R_{\mathbf nij}=\mathbf R_j+\mathbf R_\mathbf n-\mathbf R_i
\end{equation}
The infinite sum of the dipole tensor over all periodic copies of the unit cell, indexed by $\mathbf n$, and skipping the $\mathbf n=\mathbf 0$, $i=j$ terms, is evaluated efficiently with Ewald summation \citep{deLeeuwPRSLA80a},
\begin{equation}
\begin{multlined}
\mathbf{T}_{ij}(\mathbf{q})
\approx\sum_\mathbf n^{0<R_{\mathbf nij}<R_\text c}\mathbf
T^\text{erfc}(\mathbf R_{\mathbf nij},\gamma)\mathrm e^{-\mathrm i\mathbf
q\cdot\mathbf R_{\mathbf nij}} \\
+\frac{4\pi}\Omega\sum_{\mathbf
m}^{0<k_\mathbf m<k_\text c}\mathbf{\hat k}_\mathbf
m\otimes\mathbf{\hat k}_\mathbf m\,\mathrm e^{-\frac{k_\mathbf
m^2}{4\gamma^2}-\mathrm i\mathbf G_\mathbf m\cdot\mathbf R_{ij}}
\\ -\delta_{ij}\frac{4\gamma^3}{3\sqrt\pi}\mathbf I +\delta(\mathbf q)\frac{4
\pi}{3\Omega}\mathbf I,\qquad \mathbf k_\mathbf m=\mathbf G_\mathbf
m+\mathbf q
\end{multlined}\label{eq:ewald}
\end{equation}
Here, $\mathbf m$ indexes the cells of the reciprocal lattice, $\mathbf G_\mathbf m$ is a reciprocal-lattice vector, $R_\text c$ and $k_\text c$ are the real-space and reciprocal-space cutoffs, respectively, $\gamma$ is the Ewald range-separation parameter (see \cref{sec:convergence}), $\Omega$ is the unit-cell volume, and $\mathbf T^\text{erfc}$ is the short-range part of the bare dipole tensor,
\begin{equation}
\begin{aligned}
T_{ab}^\text{erfc}(\mathbf r,\gamma)
&=\frac{-3r_a r_b C(r,\gamma)+\delta_{ab}r^2 B(r,\gamma)}{r^5} \\
B(R,\gamma)
&=\operatorname{erfc}(\gamma R)
+\frac{2\gamma R}{\sqrt\pi}\mathrm e^{-{(\gamma R)}^2} \\
C(R,\gamma)
&=3\operatorname{erfc}(\gamma R)
+\frac{2\gamma R}{\sqrt\pi}(3+2{(\gamma R)}^2)\mathrm e^{-{(\gamma R)}^2}
\end{aligned}\label{eq:terfc}
\end{equation}
The last term in \cref{eq:ewald} is the so-called surface term \citep{deLeeuwPRSLA80,BalleneggerJCP14}, which does not contribute to the energy (in practice the $\mathbf q$-point mesh avoids the $\Gamma$-point) or most other physical observables, and is stated here only for completeness.

\subsection{Self-consistent screening}
\label{sec:scs}

When put in an external electric field $\mathbf E^\text{ext}$, a polarizable system partially (or fully) screens the field through induced polarization that is self-consistent with the total field.
For charged harmonic oscillators interacting via some dipole interaction tensor $\mathbf T^\text{scs}$, this reads
\begin{equation}
  \mathbf p_i=-\alpha_i\mathbf E_i=-\alpha_i\bigg(\sum_j\mathbf T^\text{scs}_{ij}\mathbf p_j+\mathbf E_i^\text{ext}\bigg)
\end{equation}
or equivalently in the shorthand notation
\begin{gather}
  \mathbf p=-\boldsymbol\alpha(\mathbf T^\text{scs}\mathbf p+\mathbf E^\text{ext})=-{\big(\boldsymbol\alpha^{-1}+\mathbf T^\text{scs}\big)}^{-1}\mathbf E^\text{ext} \\
  \bar{\boldsymbol\alpha}\equiv{\big(\boldsymbol\alpha^{-1}+\mathbf T^\text{scs}\big)}^{-1} \label{eq:scs}
\end{gather}
Here, $\bar{\boldsymbol\alpha}_{ij}$ is a nonlocal anisotropic polarizability that describes the dipole on the $i$-th oscillator induced by an external field on the $j$-th oscillator.

In the MBD@rsSCS method, self-consistent screening is used to obtain a refined set of oscillator parameters before they are used in the MBD Hamiltonian.
To do this, one can assume a homogeneous external electric field and contract $\bar{\boldsymbol\alpha}$ to effective local polarizability, and average out the angular dependence to obtain isotropic polarizability,
\begin{equation}
\bar\alpha_i=\tfrac13\operatorname{Tr}\big[\textstyle\sum_j\bar{\boldsymbol\alpha}_{ij}\big]
  \label{eq:contraction}
\end{equation}
Effective screened oscillator frequencies $\bar\omega_i$ are obtained through $C_6$ coefficients
\begin{equation}
\bar\omega=\frac{4\bar C_6}{3\bar\alpha_0^2},\qquad
\bar C_6=\frac3\pi\int_0^\infty\mathrm du\,{\bar\alpha(\mathrm iu)}^2
\label{eq:casimir-polder}
\end{equation}
by screening the dynamic oscillator polarizability evaluated at imaginary frequency,
\begin{equation}
\alpha(\mathrm iu)=\frac{\alpha_0}{1+u^2/\omega^2},\qquad \alpha(0)\equiv\alpha_0
\label{eq:dyn-pol}
\end{equation}
In practice, the isotropic local polarizability in \cref{eq:dyn-pol} is evaluated on a quadrature grid of imaginary frequencies, screened through \cref{eq:scs} at each grid point, contracted via \cref{eq:contraction}, and the integral in \cref{eq:casimir-polder} is calculated numerically (see \cref{sec:convergence}) to yield the $C_6$ coefficients and oscillator frequencies.

\subsection{Analytical gradients}

Analytical formulas for the gradients of the MBD energy with respect to all Hamiltonian input parameters can be obtained straightforwardly by application of the derivative chain rule.
The two steps related to matrix diagonalization in \cref{eq:diagonalization} and matrix inversion and contraction in \cref{eq:scs,eq:contraction}, which are nontrivial, and for which naive application of the chain rule could result in inefficient evaluation of the gradients are explicitly stated below.
When implemented, these formulas can be straightforwardly parallelized and result in a computational cost for the gradients that scales with the same power of system size as the energy evaluation.

The gradient of the MBD energy with respect to any Hamiltonian oscillator parameter $X_i$ is
\begin{equation}
\frac{\partial E_\text{MBD}}{\partial X_i}=
\operatorname{Re}\frac12\sum_{ajb}{(
\mathbf C\boldsymbol\Lambda^{-\frac12}\mathbf C^\dagger
)}_{ia,jb}
\frac{\partial Q_{jb,ia}}{\partial X_i}-
\frac32\frac{\partial\omega_i}{\partial X_i}
\label{eq:mbd-gradient}
\end{equation}
The gradient of the contracted SCS polarizability with respect to any oscillator parameter $X_i$ is
\begin{equation}
\begin{gathered}
\frac{\partial\bar\alpha_i}{\partial X_j}=
-\frac13\sum_{ab}\Big(\textstyle
B_{ia,jb}\sum_k\bar\alpha_{ka,jb}+
\bar\alpha_{jb,ia}\sum_k B_{jb,ka}
\Big) \\
\mathbf B=\boldsymbol{\bar\alpha}\mathbf A,
\qquad A_{ia,jb}=
\delta_{ij}\delta_{ab}\frac{\partial(\alpha_i^{-1})}{\partial X_i}
+\frac{\partial T^\text{scs}_{ia,jb}}{\partial X_i}
\end{gathered}
\label{eq:scs-gradient}
\end{equation}

\subsection{MBD properties}
\label{sec:mbd-properties}

The normalized ground-state wave function of the molecular MBD Hamiltonian can be expressed as
\begin{equation}
\Psi(\boldsymbol\xi)=\Bigg(\prod_{k=1}^{3N}{\Big(\frac{\tilde\omega_k}{\pi}\Big)}^\frac14\Bigg)\exp\bigg(-\frac12\boldsymbol\xi\cdot\mathbf C\boldsymbol\Lambda^\frac12\mathbf C^\dagger\boldsymbol\xi\bigg)
\end{equation}
While only two parameters per oscillator, $\alpha_{0,i}$ and $\omega_i$, are needed to get the MBD energy, a third parameter, such as $m_i$ or $q_i$, is needed to evaluate the wave function in real space, in order to convert from $\boldsymbol\xi_i$ to $\mathbf r_i$.
The wave function of the non-interacting oscillators is simply
\begin{equation}
\Psi_0(\boldsymbol\xi)=\Bigg(\prod_{i=1}^{N}{\Big(\frac{\omega_i}{\pi}\Big)}^\frac34\Bigg)\exp\bigg(-\sum_i\frac12\omega_i\xi_i^2\bigg)
\end{equation}

The MBD wave function has been used to calculate two properties of interest---the polarization of the electron density due to vdW interactions \citep{HermannNC17},
\begin{equation}
\begin{aligned}
n_\text{pol}(\mathbf r)&=\langle\Psi|\hat n|\Psi\rangle -\langle\Psi_0|\hat n|\Psi_0\rangle \\
\hat n&=\sum_i q_i\delta(\mathbf r-\mathbf r_i)
\end{aligned}
\label{eq:density}
\end{equation}
and the first-order correction to the MBD energy from the Coulomb interaction \citep{StohrNC21},
\begin{equation}
\begin{aligned}
E_\text{Coul}^{(1)}&=\langle\Psi|\hat V_\text{Coul}-\hat V_\text{dip}|\Psi\rangle-\langle\Psi_0|\hat V_\text{Coul}|\Psi_0\rangle \\
\hat V_\text{Coul}&=\frac12\sum_{i\neq j}q_i q_j\Big(\textstyle\frac1{\lvert\mathbf r_i-\mathbf r_j\rvert}
-\frac2{\lvert\mathbf r_i-\mathbf R_j\rvert}+\frac1{\lvert\mathbf R_i-\mathbf R_j\rvert}\Big) \\
\hat V_\text{dip}&=\frac12\sum_{i\neq j}q_i q_j(\mathbf r_i-\mathbf R_i)\cdot\mathbf T_{ij}(\mathbf r_j-\mathbf R_j)
\end{aligned}\label{eq:coulomb}
\end{equation}
All these expectation values can be evaluated analytically, and the derivations and final results can be found in \citet{Hermann18}.

Instead of diagonalizing the MBD Hamiltonian, the MBD energy can be equivalently obtained as an integral over the imaginary frequency, which also allows for a many-body expansion of the energy in the orders of $\mathbf T^\text{lr}$ \citep{TkatchenkoJCP13},
\begin{align}
E_\text{MBD}&=\frac1{2\pi}\int_0^\infty\mathrm du\operatorname{Tr}\big[\ln\big(\mathbf I+\boldsymbol\alpha(\mathrm iu)\mathbf T^\text{lr}\big)\big] \label{eq:log-formula} \\
&=-\frac1{2\pi}\int_0^\infty\mathrm du\sum_{n=2}^\infty\frac{{(-1)}^n}{n}\operatorname{Tr}\big[{\big(\boldsymbol\alpha(\mathrm iu)\mathbf T^\text{lr}\big)}^n\big] \label{eq:rpa}
\end{align}
In practice, $\boldsymbol\alpha\mathbf T^\text{lr}$ can be replaced with the Hermitian $\boldsymbol\alpha^{1/2}\mathbf T^\text{lr}\boldsymbol\alpha^{1/2}$ since both are equivalent under the trace operator.

This integral formulation has two applications.
First, one can analyze the contribution of the second-order (pairwise) and higher many-body orders to the vdW energy.
Second, the eigenvalues $x_k$ of $\boldsymbol\alpha(\mathrm iu)\mathbf T^\text{lr}$ can be heuristically renormalized to obtain a real-valued MBD energy even when the exact diagonalization gives a complex-valued energy due to the dipole collapse \citep{GouldJCTC16a}.
In particular, one replaces
\begin{equation}
\begin{gathered}
\ln(1+x_k):=\ln(1+\tilde x_k)-\tilde x_k \\
\tilde x_k=\begin{cases}
x_k & x_k\geq0 \\
-\operatorname{erf}{\Big(\frac{\sqrt{\pi}}2 x_k^4\Big)}^{1/4}  & x_k < 0
\end{cases}
\end{gathered}\label{eq:gould-norm}
\end{equation}

Another property of interest is the macroscopic dielectric constant $\boldsymbol\epsilon_\mathrm M$ of a periodic system of charged oscillators, which can be calculated via SCS from $\bar{\boldsymbol\alpha}$ as
\begin{equation}
\boldsymbol\epsilon_\mathrm M=\lim_{\mathbf q\rightarrow 0}{\bigg(1+\frac{4\pi}\Omega\Big(\textstyle\sum_{ij}\bar{\boldsymbol\alpha}_{ij}(\mathbf q)\Big)\bigg)}^{-1}
\end{equation}

\subsection{Pairwise dispersion}

The second (lowest) order of the many-body expansion in \cref{eq:rpa} corresponds to the familiar pairwise expression for the vdW energy,
\begin{equation}
\begin{gathered}
E_\text{MBD}^{(2)}=-\frac12\sum_{i\neq j}C_{6,ij}\frac{f(\boldsymbol\kappa_i,\boldsymbol\kappa_j,R_{ij})}{R_{ij}^6} \\
C_{6,ij}=\frac32\frac{\alpha_{0,i}\alpha_{0,j}}{\omega_i^{-1}+\omega_j^{-1}}, \quad
f(\boldsymbol\kappa_i,\boldsymbol\kappa_j,R_{ij})=\frac{R_{ij}^6}6\operatorname{Tr}\big[{(\mathbf T_{ij}^\text{lr})}^2\big]
\end{gathered}
\label{eq:pairwise}
\end{equation}
Here, $f$ is a damping function ($f\rightarrow 1$ when $R_{ij}\rightarrow\infty$) and is typically specified directly rather than through $\mathbf T^\text{lr}$.
This \textit{ansatz} has been used in numerous vdW models, among them the predecessor of MBD, the TS method \citep{TkatchenkoPRL09}.

Under periodic boundary conditions, the corresponding lattice sum of \cref{eq:pairwise} converges absolutely, unlike the dipole tensor, but the rate of convergence can be slow enough to present a computational bottleneck.
An alternative is to use the Ewald technique \citep{KarasawaJPC89},
\begin{equation}
\begin{gathered}
  \begin{multlined}
  \sideset{}{'}\sum_{\mathbf n}\frac1{R_{\mathbf nij}^6}
  \approx\sum_\mathbf n^{0<R_{\mathbf nij}<R_\text c}\gamma^6\phi_r(\gamma R_{\mathbf nij})
  \\
  +\frac1\Omega\sum_{\mathbf
  m}^{0<G_\mathbf m<k_\text c}\gamma^3\phi_k\left(\frac{G_\mathbf
  m}\gamma\right)\cos(-\mathrm i\mathbf G_\mathbf m\cdot\mathbf R_{ij})+\delta_{ij}\frac{\gamma^6}{6}
  \end{multlined} \\
  \begin{aligned}
  \phi_r(r)&=\Big(\textstyle\frac1{r^6}+\frac1{r^4}+\frac1{2r^2}\Big)\mathrm e^{-r^2} \\
  \phi_k(k)&=\frac{\pi^\frac32}{12}\Big((4-2k^2)\mathrm e^{-\frac{k^2}4}+k^3\sqrt\pi\operatorname{erfc}\big(\tfrac12 k\big)\Big)
  \end{aligned}
\end{gathered}\label{eq:ts-ewald}
\end{equation}
with the same notation as in \cref{eq:ewald}.

\subsection{MBD methods}
\label{sec:mbd-methods}

The general MBD framework discussed up to this point is concretized into an MBD method by specifying at least two aspects.
First, how are the MBD oscillators parametrized ($\alpha_0$, $\omega$).
Second, how is $\mathbf T^\text{lr}$ chosen to smoothly integrate the long-range correlation from MBD with a given method that accounts for the short-range correlation, typically KS-DFT or its substitute.
While independent to a certain degree, these two choices are ultimately related in practice.

\paragraph{TS}

In the pairwise TS method \citep{TkatchenkoPRL09}, each atom is mapped to a single oscillator parametrized through scaling of reference free-atom vdW parameters based on atomic volumes,
\begin{equation}
\begin{aligned}
  \alpha_{0,i}&:=\alpha_{0,i}^\text{ref}\gamma_i,\qquad \gamma_i=\frac{V_i}{V_i^\text{ref}} \\
  \omega_i&:=\frac{4C_{6,ii}^\text{ref}\gamma_i^2}{3{\big(\alpha_{0,i}^\text{ref}\gamma_i\big)}^2}=\frac{4C_{6,ii}^\text{ref}}{3{\big(\alpha_{0,i}^\text{ref}\big)}^2}=\omega_i^\text{ref} \\
  R_i^\text{vdW}&:=R_i^\text{vdW,ref}{\gamma_i}^\frac13
\end{aligned}\label{eq:ts-scaling}
\end{equation}
Here, $V_i$ is some measure of a volume of an atom in a molecule or material and $V_i^\text{ref}$ is the same measure for a corresponding free atom.
Possible choices for the volume measure are discussed in \cref{sec:integration}.

The logistic sigmoid function parametrized with volume-rescaled vdW radii $R^\text{vdW}$ is used as the damping function in \cref{eq:pairwise},
\begin{equation}
\begin{gathered}
  f(R_i^\text{vdW},R_j^\text{vdW},R_{ij}):={\big(1+\mathrm e^{-a(\eta-1)}\big)}^{-1} \\
  \eta={\textstyle\frac{R_{ij}}{s_R(R_i^\text{vdW}+R_j^\text{vdW})}}
\end{gathered}\label{eq:fermi-damping}
\end{equation}
The damping parameter $s_R$ adjusts the onset of the vdW interaction and is optimized separately for each individual short-range correlation model, and $a=20$.

\paragraph{MBD@rsSCS}

The MBD@rsSCS method \citep{AmbrosettiJCP14} is the many-body extension of the TS method, in which the oscillators interact through the dipole tensor derived from the full Coulomb interaction of two oscillator (Gaussian) charge densities with widths $\sigma_i^2$,
\begin{equation}
\begin{gathered}
\begin{aligned}
T^\text{GG}_{ab}(\mathbf r,\sigma)&=
\frac{\partial^2}{\partial r_a\partial r_b}\frac{\operatorname{erf}(\zeta)}r \\
&=\big(\operatorname{erf}(\zeta)-\Theta(\zeta)\big)T_{ab}(\mathbf r)+
2\zeta^2\Theta(\zeta)\frac{r_a r_b}{r^5}
\end{aligned} \\
\Theta(\zeta)=\frac{2\zeta}{\sqrt\pi}\mathrm e^{-\zeta^2},\qquad
\zeta=\frac r{\sqrt{\sigma_i^2+\sigma_j^2}}
\end{gathered}\label{eq:tgg}
\end{equation}
The oscillator width is related to its polarizability through certain self-consistence requirements \citep{MayerPRB07},
\begin{equation}
\sigma_i={\bigg(\frac13\sqrt{\frac2\pi}\alpha_i\bigg)}^{\frac13}
\label{eq:mayer}
\end{equation}
To integrate this full-range interaction with KS-DFT while avoiding double-counting of the short-range correlation, $\mathbf T^\mathrm{GG}$ is range-separated, and the short-range part is used in SCS to refine the TS oscillator parameters from \cref{eq:ts-scaling}, while the long-range part is used in the MBD Hamiltonian,
\begin{align}
\mathbf T^\text{scs}&:=\big(1-g(R_i^\text{vdW},R_j^\text{vdW},R_{ij})\big)\mathbf T^\mathrm{GG}(\alpha_i,\alpha_j,\mathbf R_{ij}) \label{eq:scs-sr-dipole}\\
\mathbf T^\text{lr}&:=g(\bar R_i^\text{vdW},\bar R_j^\text{vdW},R_{ij})\mathbf T^\mathrm{GG}(\alpha_i,\alpha_j,\mathbf R_{ij}) \notag\\
&\approx g(\bar R_i^\text{vdW},\bar R_j^\text{vdW},R_{ij})\mathbf T(\mathbf R_{ij}) \label{eq:mbd-lr-dipole} \\
&\qquad\bar R_\text{vdW}=R_\text{vdW}{\Big(\frac{\bar\alpha_0}{\alpha_0}\Big)}^\frac13 \label{eq:scale-rvdw-scs}
\end{align}
Here, $g$ is a damping function of the same form as $f$ in \cref{eq:fermi-damping}, but with $a=6$ and $s_R$ usually denoted as $\beta$.

\paragraph{MBD-NL}

The MBD-NL method \citep{HermannPRL20} removes the need for SCS, and the MBD Hamiltonian is parametrized directly with
\begin{equation}
  \alpha_{0,i}:=\alpha_{0,i}^\text{ref}\frac{\alpha_{0,i}^\text{df}}{\alpha_{0,i}^\text{df,ref}},\qquad
  \omega_i:=\omega_i^\text{ref}\frac{\omega_i^\text{df}}{\omega_i^\text{df,ref}}
  \label{eq:mbd-nl-scaling}
\end{equation}
Here, a polarizability functional of the density is used to obtain vdW parameters ($\alpha_0^\text{df}$, $\omega^\text{df}$) for an atom in a molecule or material and the corresponding free atom (see \cref{sec:integration}).
MBD-NL uses the same parametrization of $\mathbf T^\text{lr}$ in \cref{eq:mbd-lr-dipole} as MBD@rsSCS, but with vdW radii in \cref{eq:fermi-damping} replaced with
\begin{equation}
  R_i^\text{vdW}:=\frac52{(\alpha_{0,i}^\text{ref})}^\frac17{\Bigg(\frac{\alpha_{0,i}^\text{df}}{\alpha_{0,i}^\text{df,ref}}\Bigg)}^\frac13
  \label{eq:mbd-nl-rvdw}
\end{equation}

\subsection{Integration with short-range models}
\label{sec:integration}

MBD models the long-range part of electron correlation and must be always coupled with a model that captures the rest of the electronic energy, including the short-range correlation,
\begin{equation}
E=E_\text{sr}+E_\text{MBD}
\end{equation}
Typically, this short-range model is some semilocal or hybrid version of KS-DFT, but it can also be its computationally more efficient substitute such as DFTB or a machine-learned interatomic potential.
In any case, the short-range model and MBD must be integrated in two ways.
First, the damping parameter in an MBD method (\cref{sec:mbd-methods}) must be adjusted to avoid double counting of correlation in the intermediate range.
This is typically done by fitting on the S66x8 dataset of vdW dimers of small organic molecules \citep{BrauerPCCP16}.
Second, the short-range model is used to parametrize the MBD Hamiltonian, that is, to obtain the vdW parameters of each oscillator.
How this is done depends on a particular model and the various options are described below.
In the context of libMBD, the following equations must be necessarily evaluated in the code that implements the short-range model and couples to libMBD.

\paragraph{Hirshfeld volumes}

This parametrization uses \cref{eq:ts-scaling}, in which the atomic volumes are calculated from the Hirshfeld-partitioned electron density obtained from a DFT calculation,
\begin{equation}
\begin{gathered}
  V_i:=V[n_i],\qquad V_i^\text{ref}:=V[n_i^\text{ref}] \\
  V[n]=\int\mathrm d\mathbf r n(\mathbf r)r^3,\qquad n_i(\mathbf r)=n(\mathbf R_i+\mathbf r)w_i^\text H(\mathbf r) \\
  w_i^\text H(\mathbf r)=\frac{n_i^\text{ref}(\mathbf r)}{\sum_j n_j^\text{ref}(\mathbf r+\mathbf R_i-\mathbf R_j)}
\end{gathered}\label{eq:hirshfeld-vol}
\end{equation}
Here $n_i^\text{ref}$ are electron densities of free atoms located at $\mathbf r=0$.
A straightforward extension is to use not only neutral free atoms, but also free ions \citep{BuckoJCP14} for the density partitioning, which improves description of ionic compounds.
This was further extended to use not only ion densities, but also ionic reference vdW parameters by using a piecewise linear dependence of atomic polarizabilities on the charge \citep{GouldJCTC16a}.
This results in atomic $\alpha_i(\mathrm iu)$ that cannot be expressed through a single oscillator with \cref{eq:dyn-pol}, requiring either direct use of \cref{eq:log-formula} or calculation of effective oscillator frequencies via \cref{eq:casimir-polder}.

\paragraph{MBD-NL}

Electron density from a KS-DFT calculation is also used in MBD-NL to parametrize the MBD Hamiltonian.
Here, a renormalized VV10 polarizability functional of the density \citep{VydrovJCP10a} is coarse-grained via Hirshfeld partitioning,
\begin{equation}
  \alpha_i^\text{df}(\mathrm iu)
    :=\int\mathrm d\mathbf r\frac{g(I,\chi)n(\mathbf r)w_i^\text{H}(\mathbf r)}{\frac{4\pi}3n(\mathbf r)
    +C\frac{{|\boldsymbol\nabla n(\mathbf r)|}^4}{n{(\mathbf r)}^4}+u^2}
\end{equation}
and the resulting atomic dynamic polarizabilities are reduced to effective oscillator frequencies ($\omega_i^\text{df}$) via \cref{eq:casimir-polder}.
The free-atom reference values ($\alpha_{0,i}^\text{df,ref}$, $\omega_i^\text{df,ref}$) are obtained by evaluating the functional on free-atom densities.
The renormalized VV10 functional uses a semilocal cutoff function $g$ \citep[see][Eq.~7]{HermannPRL20}, which is a function of the local ionization potential $I$ and the iso-orbital indicator $\chi$, which in turn are functions of the electron density, its gradient, and the kinetic energy density, quantities readily available in any code that implements meta-GGA functionals.

\paragraph{Charge population analysis}

In the absence of a real-space representation of the electron density, the above-mentioned methods are not applicable.
Many electronic structure methods exist where the electron density is not at all or only rarely evaluated explicitly.
This includes approximate methods such as DFTB, as well as other density-matrix-based approaches.
\citet{StohrJCP16} found that a correlation between effective atomic polarizability of an atom in a molecule can be established based not only on the effective volume of an atom, but also on measures derived from the density matrix represented in a local atomic orbital basis set, $\lvert\psi_a\rangle = \sum_{\nu} c_{\nu}^a \lvert\phi_{\nu}\rangle$, where single-electron states $\lvert\psi_a\rangle$ are expanded in atomic orbital basis functions $\lvert\phi_{\nu}\rangle$.
When scaling free-atom vdW parameters (\pcref{eq:ts-scaling}) with the degree of hybridization between atoms as measured by the sum over the on-site component of the Mulliken charges (or equivalently the atom-projected trace over the density matrix) and the atomic charge $Z_i$,
\begin{equation}
    \gamma_i:=\frac{h_i}{Z_i} = \frac{\sum_a f_a \sum_{\nu \in i}{|c_{\nu}^a|}^2}{Z_i}
\end{equation}
static atomic polarizabilities can be predicted that are similar to the ones obtained by rescaling with Hirshfeld volumes (\pcref{eq:hirshfeld-vol}).

\paragraph{Machine-learned interatomic potentials}

Several works have shown that the effective atoms-in-molecules volume scaling ratios as a function of the atomic positions can be accurately represented by machine-learning regression \citep{BereauJCP18,WestermayrDD22,PoierJPCL22}.
Once a continuous representation of the scaling ratio is constructed, an MBD description of the long-range correlation energy can be straightforwardly coupled to short-range force fields or atomistic machine-learned potentials.
In cases where these potentials are directly trained on semilocal approximations to DFT, the choice of the range-separation parameter remains identical to the optimal choice for the underlying density-functional approximation.

\section{Implementation}
\label{sec:implementation}

\subsection{Code structure and API}

\begin{figure}[t]
\centering
\includegraphics{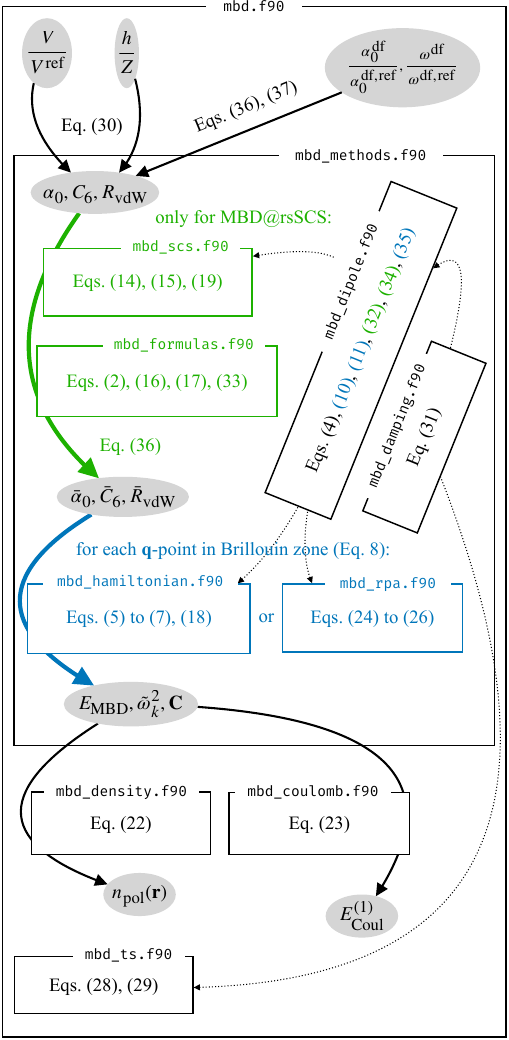}
\caption{\textbf{Mapping code to physics.}
libMBD was designed for ease of development, which is achieved through tight correspondence between code and physics.
An overview of an MBD@rsSCS calculation is shown on the background of a map of MBD source files with the individual implemented equations.
}\label{fig:libMBD}
\end{figure}

\begin{lstlisting}[language=Fortran,float=t,caption={libMBD's Fortran API minimal example.},label=lst:fortran-api]
use mbd, only: mbd_input_t, mbd_calc_t
use iso_fortran_env, only: real64

type(mbd_input_t) :: inp
type(mbd_calc_t) :: calc
real(real64) :: energy, gradients(3, 2)
integer :: code
character(200) :: origin, msg

inp%atom_types = ['Ar', 'Ar']
inp%coords = reshape([0d0, 0d0, 0d0, 0d0, 0d0, 7.5d0], [3, 2])
inp%method = 'mbd-rsscs'
inp%xc = 'pbe'
call calc%init(inp)
call calc%get_exception(code, origin, msg)
if (code > 0) then
    print *, msg
    stop 1
end if
call calc%update_vdw_params_from_ratios([0.98d0, 0.98d0])
call calc%evaluate_vdw_method(energy)
call calc%get_gradients(gradients)
call calc%destroy()
\end{lstlisting}

\begin{lstlisting}[language=Python,float=t,caption={libMBD's Python API minimal example.},label=lst:python-api]
from pymbd.fortran import MBDGeom

geom = MBDGeom([(0, 0, 0), (0, 0, 7.5)])
energy = geom.mbd_energy_species(
    ['Ar', 'Ar'], [0.98, 0.98], beta=0.83
)
\end{lstlisting}


libMBD implements all equations in \crefrange{sec:mbd}{sec:mbd-methods} in Fortran 2008.
To enable straightforward and rapid development of new variants and extensions of MBD, libMBD has been designed from the start to maintain a close correspondence between code and physics (\cref{fig:libMBD}).
Equations are documented in source code with \LaTeX\ right next to the individual procedures that implement them and rendered in auto-generated documentation (\url{https://libmbd.github.io}).
Gradients of the energy are computed in a forward mode, where each individual function that computes a value from its arguments also optionally returns the partial derivatives of the output value with respect to some (or all) of its input arguments.
Parallelization is available either by trivial distribution of the $\mathbf q$-point mesh (\pcref{eq:brillouin-int}) or by distribution of atom-indexed matrices ($\mathbf T$, $\mathbf Q$, $\bar{\boldsymbol\alpha}$) via \citet{BLACS}.
Linear-algebra operations are handled via \citet{LAPACK}, \citet{ScaLAPACK}, or \citet{ELPA} \citep{MarekJPCM14} interfaced through \citet{ELSI} \citep{YuCPC20}.
libMBD can be used natively from Fortran, as a plain binary library via C API made by the \verb+iso_c_binding+ Fortran module, or from Python via a C extension that internally binds to the C API\@.
The Fortran API (\cref{lst:fortran-api}) is based on two derived types, \verb+mbd_input_t+ and \verb+mbd_calc_t+, where the former is a plain data structure of all input parameters used to initialize the latter, which is an opaque handle that provides access to computation and results.
The Python API (\cref{lst:python-api}) is based on the \verb+MBDGeom+ class that represents a given geometry and provides access to computation via its methods.

80\% of the code base of libMBD is covered by tests, which include both unit tests of analytical gradients of individual procedures as well as regression tests of the total energies and gradients.
libMBD is built with \citet{CMake}.

\subsection{Scalability}

\begin{figure}[t]
\centering
\includegraphics{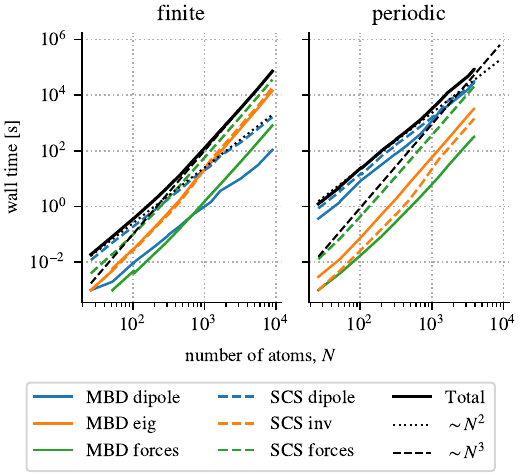}
\caption{\textbf{System-size scaling.}
The scaling of an MBD calculation with system size transitions from quadratic for small systems where the matrix construction dominates to cubic for large systems where matrix operations (multiplication, diagonalization, inversion) dominate.
}\label{fig:system-size-scaling}
\end{figure}

\begin{figure}[t]
\centering
\begin{tikzpicture}
\node[below right, inner sep=0pt] at (2.0,0.0) {\includegraphics[width=2in]{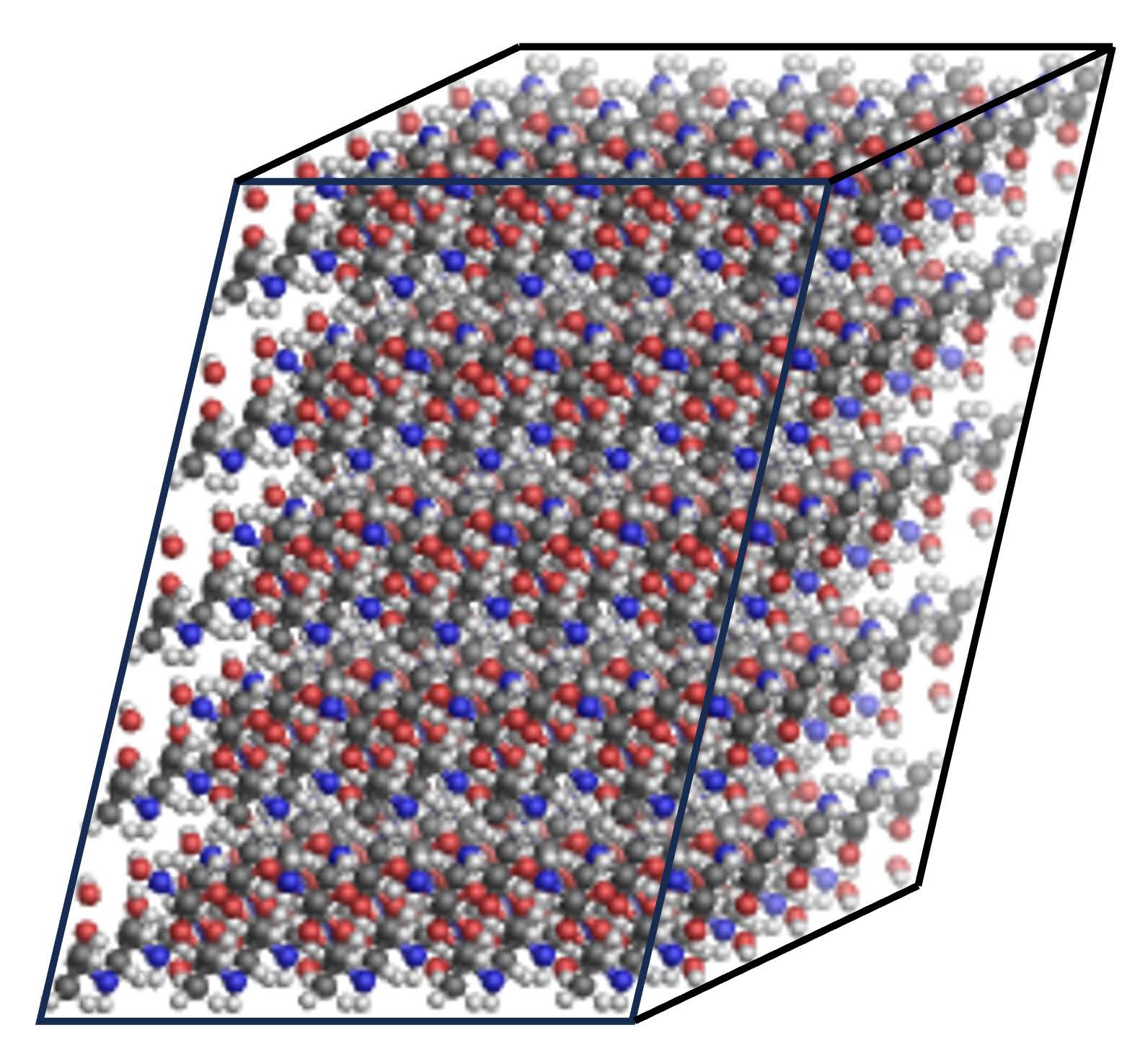}};
\node[below right] at (2.5,0.0) {\bfseries a};
\node[below right, inner sep=0pt] at (0.0,-5) {\includegraphics{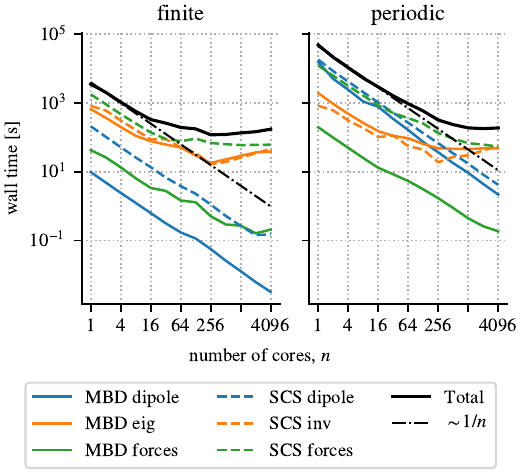}};
\node[below right] at (0,-5) {\bfseries b};
\end{tikzpicture}
\caption{\textbf{Strong scaling.}
(\textbf a) $5\times5\times5$ urethane supercell used for benchmarking.
(\textbf b) libMBD parallelizes well to hundreds of CPU cores, and is mostly limited only by the parallelization of the matrix operations in external libraries.
The current simple implementation of the SCS forces uses an unnecessary amount of inter-process communication, resulting in a relatively early plateau for finite systems.
This will be addressed in future work.
}\label{fig:strong-scaling}
\end{figure}

For libMBD to be practical, it needs to keep up with the codes that implement the corresponding short-range models (DFT and its substitutes) in terms of scalability to large systems, and scalability to a large number of processors.
libMBD uses only process parallelization through the MPI and BLACS, and no thread parallelization.

Tested on various supercells of crystalline urethane (unit cell with 26 atoms), with or without periodic boundary conditions, the MBD method and its implementation in libMBD scale with the number of atoms, $N$, as follows (\cref{fig:system-size-scaling}).
For small $N$, the computational cost is dominated by the construction of the various matrices and as such grows quadratically with $N$.
At around a few hundred atoms (finite system) or a few thousand atoms (periodic system), the matrix operations that scale cubically with $N$ (multiplication, diagonalization, inversion) start to dominate.
The MBD analytical gradients as implemented by \citet{Blood-ForsytheCS16} add one extra factor of $N$ to the scaling (making it quartic) due to explicit evaluation of the gradient of $\bar{\boldsymbol\alpha}$ (see Eqs.~44 and~45 therein).
libMBD avoids this by calculating only the gradient of the fused evaluation of $\bar{\boldsymbol\alpha}$ and its contraction to atoms (\pcref{eq:scs-gradient}), and as a result calculation of the gradients changes only the prefactor in the scaling behavior.

Tested on a $5\times5\times5$ supercell of urethane (3,250 atoms), libMBD exhibits good strong scaling with the number of processor cores $n$ (\cref{fig:strong-scaling}).
The construction of the various matrices ($\mathbf T$, $\mathbf Q$, $\bar{\boldsymbol\alpha}$) is trivially parallelizable and has perfect strong scaling up to $n=4096$.
The evaluation of the MBD gradients requires only very little inter-process communication and shows similarly good behavior.
The distributed matrix inversion (SCS) and diagonalization (MBD) as provided by ScaLAPACK and ELPA exhibits strong scaling up to $n=256$ on our particular setup, at which point the scaling plateaus.
The evaluation of SCS gradients requires only matrix multiplications and contractions, but its current simple implementation in libMBD requires an unnecessary amount of inter-process communication, making the strong scaling plateau already at $n=64$ for finite systems.
Note that this slight inefficiency still allows the calculation of the MBD energy and gradients in $\sim$\SI{100}{\second} at $n=256$, and furthermore it affects only the MBD@rsSCS method, not the newer MBD-NL method which does not use SCS\@.
Nevertheless if better strong scaling is required in the future, the inter-process communication in the evaluation of the SCS gradients can be massively optimized.

\subsection{Convergence and default parameters}
\label{sec:convergence}

\begin{figure}[t]
\centering
\includegraphics{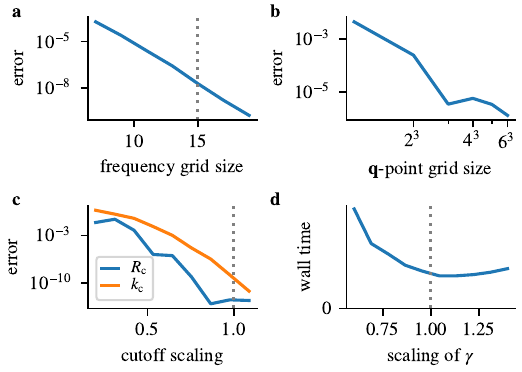}
\caption{\textbf{Numerical convergence.}
Where possible, libMBD offers sane defaults for parameters that ensure numerical convergence.
All errors are relative.
Dotted vertical line denotes the default value of a parameter. (\textbf a) The MBD energy converges exponentially with respect to the frequency grid (\pcref{eq:casimir-polder}). (\textbf b) Polynomial convergence with respect to the size of the $\mathbf q$-point grid (\pcref{eq:brillouin-int}). (\textbf c) Exponential convergence with respect to the real- and reciprocal-space cutoffs, $R_\mathrm c$, $k_\mathrm c$, in the Ewald summation (\pcref{eq:ewald}). (\textbf d) Optimal range separation in Ewald summation to minimize computational cost.}
\label{fig:convergence}
\end{figure}

There are a few critical parameters that determine the numerical accuracy of the calculated MBD energy with respect to the exact energy as defined in \crefrange{sec:mbd}{sec:mbd-methods}.
The SCS is evaluated and integrated on an imaginary-frequency grid (\pcref{eq:casimir-polder}), for which libMBD uses a Gauss--Legendre quadrature transformed from $[-1,1]$ to $[0,\infty]$ with $x\rightarrow0.6\cdot(1+x)/(1-x)$.
The MBD energy converges exponentially with the grid size, and using the default 15 grid points, it is converged to relative error of $10^{-8}$ (\cref{fig:convergence}a).
The frequency grid size is the only parameter for calculations on finite systems.
For periodic systems, there are additional parameters related to converging the infinite lattice sums involved.

The MBD energy per unit cell of a periodic system is calculated by integration over the FBZ (\pcref{eq:brillouin-int}).
libMBD uses a uniform quadrature grid in the reciprocal space, offset from the origin since the MBD energy has a removable singularity at the $\Gamma$-point (see \cref{sec:pbc}).
The convergence of the MBD energy with the $\mathbf q$-point grid size is polynomial and system-dependent (see \cref{fig:convergence}b for example convergence on urethane), and as such the grid needs to be specified by the user.

The Ewald summation of the dipole interactions between the origin cell and the infinite crystal (\pcref{eq:ewald}) is controlled by three interconnected parameters---the range-separation parameter $\gamma$, and the real- and reciprocal-space cutoffs $R_\mathrm c$, $k_\mathrm c$.
libMBD uses the following default prescription for the parameters,
\begin{equation}
R_\text c:=\frac6\gamma,\quad k_\text c:=10\gamma, \quad \gamma:=\frac{2.5}{\sqrt[3]\Omega}
\end{equation}
where $\Omega$ is the unit-cell volume.
The MBD energy converges exponentially with the cutoffs and the default values ensure convergence to relative error of $10^{-10}$ (\cref{fig:convergence}c).
The range-separation parameter does not affect accuracy, but through its determination of the cutoffs affects computational cost.
The default value in libMBD ensures an optimal balance between the costs of the real- and reciprocal-space parts (\cref{fig:convergence}d).

\subsection{Functionality}
\label{sec:functionality}

\begin{table}
\caption{Available features in libMBD.}
\label{tab:functionality}
\centering
\setlength{\tabcolsep}{3pt}
\begin{tabularx}{\linewidth}{>{\raggedright\arraybackslash}Xcccc}
\toprule
                                & \makecell{Python\\bindings} & periodic & parallel  & gradients$^a$ \\
\midrule
MBD energy                      & ✓                           & ✓         & ✓        & ✓             \\
SCS                             & ✓                           & ✓         & ✓        & ✓             \\
TS energy                       & ✓                           & ✓         & ✓        & ✓             \\
MBD energy via $\int\mathrm du$ & ✓                           & ✓         & ✓        &               \\
MBD eigenstates                 & ✓                           & ✓         & ✓        &               \\
MBD pol.\ density               & ✓                           &           &          &               \\
MBD Coulomb corr.               & ✓                           &           &          &               \\
\bottomrule
\end{tabularx}

\footnotesize
\begin{minipage}{0.95\linewidth}
$^a$Includes forces, stress tensor, and energy gradients with respect to the vdW parameters (for self-consistence).
\end{minipage}
\end{table}

libMBD implements all equations in \crefrange{sec:mbd}{sec:mbd-methods}, with all physical quantities accessible via the Python bindings (\cref{tab:functionality}).
Namely, this includes the calculation of MBD and TS energies, evaluation of SCS, evaluation of the MBD energy via the imaginary-frequency integration (\pcref{eq:log-formula}), access to MBD eigenstates, and evaluation of the MBD polarization density and Coulomb correction.
All energy evaluations are implemented for both finite and periodic systems, parallelized, and with available analytical gradients (except for the imaginary-frequency integration).
Evaluation of MBD properties such as the Coulomb correction or density polarization is currently implemented only for finite systems and is not parallelized.

\subsection{Existing interfaces}

libMBD was originally developed alongside \citet{FHI-aims} \citep{BlumCPC09} and was tightly integrated into it.
Due to specific build requirements of FHI-aims, libMBD is currently distributed with FHI-aims in a lightly pre-processed form.
The current API of libMBD was then based on the preexisting framework for interfacing external libraries in \citet{DFTB+} \citep{HourahineJCP20}, which was the first third-party program in which libMBD was integrated.
As a result, the DFTB+ integration is very lightweight and can serve as a model for interfaces with other programs.
libMBD has been since integrated into \citet{Quantum-ESPRESSO} \citep{GiannozziJCP20}, VASP \citep{KressePRB96}, and \citet{Q-Chem} \citep{EpifanovskyJCP21} without having to change anything in the existing source code, demonstrating its universality.

While libMBD is typically integrated into a larger electronic structure program as part of a DFT or similar calculation, this is not the only use case.
Especially when developing new methods or when investigating unexpected behavior of existing methods, it is practical to be able to evaluate MBD separately from DFT, and to be able to do so as flexibly as possible.
For this reason, libMBD was from the start developed together with a Python interface called pyMBD.
Beyond using pyMBD directly (\cref{lst:python-api}), it can also serve as a bridge between libMBD and other Python programs.
One such existing example is the interface to \citet{ASE} \citep{WestermayrDD22}.

\section{Applications}
\label{sec:applications}

To demonstrate the functionality of libMBD, here we report on novel calculations done on supramolecular complexes and review selected previous applications of MBD.

\subsection{Interaction energies}

We apply the MBD scheme via the libMBD interface within the FHI-aims \citep{BlumCPC09} software package to benchmark the interaction energies of structures in the L7 dataset \citep{SedlakJCTC13} against existing data in the literature.  
We compare the interaction energies of the L7 complexes as calculated with MBD against higher-level theories, in particular diffusion Monte Carlo (DMC) \citep{ReynoldsJCP82, FoulkesRMP01} data from \citet{Al-HamdaniNC21} and \citet{BenaliJCP20a} and local natural orbital coupled cluster with single, double, and perturbative triple excitations [LNO-CCSD(T)] \citep{RolikJCP11, RolikJCP13} data from \citet{BallesterosJCP21}.
Both DMC and LNO-CCSD(T) have been shown to accurately predict intermolecular interactions of small organic molecules.
We also compare MBD-calculated interaction energies against DFT-D3 \citep{GrimmeJCP10} data from \citet{XuJCP21} and DFT-D4 data from \citet{CaldeweyherJCP17,CaldeweyherJCP19}.
Interaction energies were calculated as the difference between the total energy of the structure and the sum of the monomer energies.

DFT calculations were performed with the FHI-aims software package, using the PBE \citep{PerdewPRL96} and HSE06 \citep{HeydJCP03, KrukauJCP06} density-functional approximations.
The standard screening parameter of \SI{0.11}{\per\bohr} was used for HSE06, and the numeric atomic orbitals were represented using a `tight' basis set \citep{BlumCPC09, HavuJCP09}.
The total energy, sum of eigenvalues, and charge density criteria were set to \SI{1E-6}{\electronvolt}, \SI{1E-2}{\electronvolt}, and \SI{1E-5}{\electronvolt\per\bohr\cubed}, respectively.
Input and output files for all DFT+MBD calculations are available as a dataset in the NOMAD electronic structure data repository \citep{NOMAD_Upload}. 

\begin{figure*}[t]
\centering
\includegraphics[width=5.5in]{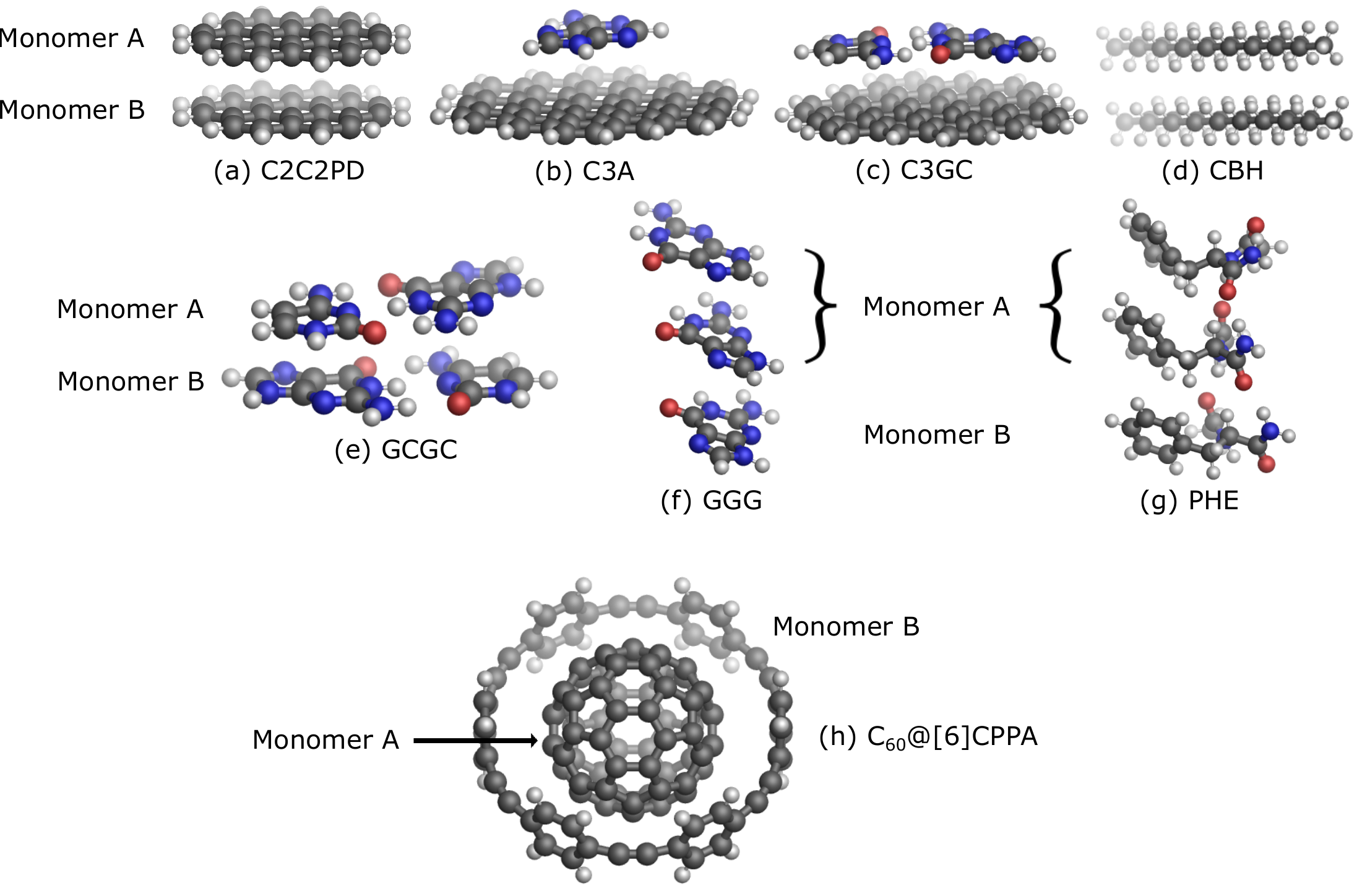}
\caption{\textbf{Supramolecular complexes.}
All complexes were split into two interacting fragments, denoted as `Monomer A' and `Monomer B'.
H, C, N and O atoms are shown in white, gray, blue, and red, respectively.
(\textbf a--\textbf g) L7 dataset.
(\textbf h) C$_{60}$@[6]CPPA complex.
}\label{fig:L7}
\end{figure*}

The L7 dataset \citep{SedlakJCTC13} was chosen as it contains non-covalent complexes considerably larger than other datasets such as S22 \citep{JureckaPCCP06} or S66 \citep{RezacJCTC11, RezacJCTC14}.
The L7 structures (\cref{fig:L7}a--g) comprise intermolecular complexes between 48 and 112 atoms in size and mostly dispersion-dominated, which makes them not only a good representation of biochemical structures but also provides a suitable test case to assess the accuracy of MBD for medium to large complexes involving $\uppi$--$\uppi$ stacking, electrostatic interactions, and hydrogen bonding \citep{Al-HamdaniNC21}.
The L7 dataset contains a parallel-displaced $\uppi$--$\uppi$-stacked coronene dimer (C2C2PD), a $\uppi$--$\uppi$-stacked coronene$\cdots$adenine dimer (C3A), a $\uppi$--$\uppi$-stacked circumcoronene$\cdots$Watson-Crick hydrogen-bonded guanine-cytosine dimer (C3GC), an octodecane dimer in stacked parallel conformation (CBH), a $\uppi$--$\uppi$-stacked Watson-Crick hydrogen-bonded guanine-cytosine dimer (GCGC), a stacked guanine trimer (GGG), and a phenylalanine residues trimer in mixed hydrogen-bonded-stacked conformation (PHE).
In addition, a larger system of a C$_{60}$ buckyball inside a [6]-cycloparaphenyleneacetylene ring (C$_{60}$@[6]CPPA), symmetrized to the $D_{3d}$ point group and comprising 132 atoms, was considered (\cref{fig:L7}h).
The C$_{60}$@[6]CPPA complex has a large polarizability \citep{AntoineJCP99} which gives rise to considerable dispersion interactions, and the confinement between the ring and the buckyball can result in non-trivial long-range repulsive interactions \citep{SadukhanPRL17, StohrNC21}.

\begin{table*}
    \centering
    \caption{Interaction energies (\si{\kcal\per\mole}) of supramolecular complexes with different methods.
}\label{tab:L7_Interaction}
    \newcommand\thd[1]{\multicolumn1c{#1}}
    \begin{tabular}{c*3{D..{3.5}}*5{D..{3.2}}}
        \toprule
        complex & \thd{\makecell{LNO-\\CCSD(T)$^{ab}$}} & \thd{FN-DMC$^{ac}$} & \thd{DMC$^d$} & \thd{\makecell{PBE\\+D3$^e$}} & \thd{\makecell{PBE0\\+D4$^a$}} & \thd{\makecell{PBE\\+MBD}} & \thd{\makecell{PBE0\\+MBD$^a$}} & \thd{\makecell{HSE06\\+MBD}} \\
        \midrule
        C2C2PD & -20.6(6) & -18.1(8) & -17.5(7) & -15.78 & -21.15 & -16.57 & -17.04 & -17.34 \\
        C3A & -16.5(8) & -15.0(1.0) & -16.6(9) & -13.58 & -16.46 & -13.43 & -13.86 & -14.05 \\
        C3GC & -28.7(1.0) & -24.2(1.3) & -25.1(9) & -22.87 & -27.60 & -22.80 & -23.48 & -23.88 \\
        CBH & -11.0(2) & -11.4(8) & -10.9(8) & -14.16 & -11.68 & -13.95 & -12.70 & -12.95 \\
        GCGC & -13.6(4) & -12.4(8) & -10.6(6) & -11.75 & -15.54 & -11.74 & -11.61 & -11.78 \\
        GGG & -2.1(2) & -1.5(6) & -2.0(4) & -1.39 & -2.10 & -1.66 & -1.27 & -1.31 \\
        PHE & -25.4(2) & -26.5(1.3) & -24.9(0.6) & -25.00 & -26.96 & -26.56 & -25.83 & -27.65 \\ 
        C$_{60}$@$[6]$CPPA & -41.7(1.7) & -31.1(1.4) & - & - & -39.47 & -29.59 & -32.14 & -33.89 \\
        \bottomrule
    \end{tabular}
\footnotesize

\begin{minipage}{0.85\linewidth}
$^a$Data from \citet{Al-HamdaniNC21}.
$^b$The indicated errors are extrapolated from the convergence of basis sets and local approximations.
$^c$The indicated errors account for the stochastic uncertainty of the estimation and identify a 95\% confidence interval (i.e.\ two standard deviations).
$^d$Data from \citet{BenaliJCP20a}.
$^e$Data from \citet{XuJCP21}.
\end{minipage}
\end{table*}

DFT+MBD performs fairly well for the supramolecular complexes with respect to the higher-level LNO-CCSD(T) and DMC methods (\cref{tab:L7_Interaction}).
For the majority of L7 complexes (C3A, C3GC, GCGC, GGG, and PHE), DFT+MBD is within chemical accuracy ($\pm$\SI{1}{\kcal\per\mole}) of LNO-CCSD(T) and/or DMC.
For complexes where the difference between dispersion-corrected DFT and higher-level interaction energies is greater than \SI{1}{\kcal\per\mole}, both PBE+MBD and HSE06+MBD generally outperform PBE+D3 with respect to the higher-level methods, as can be seen for both C2C2PD and CBH.

LNO-CCSD(T) and FN-DMC are in close agreement for five of the L7 structures (\cref{tab:L7_Interaction}).
However, their interaction energies differ by \SI{1.1}{\kcal\per\mole} for C2C2PD, \SI{2.2}{\kcal\per\mole} for C3GC, and \SI{7.6}{\kcal\per\mole} for the C$_{60}$@[6]CPPA complex.
Interestingly, PBE0+D4 is in close agreement with LNO-CCSD(T) across all complexes, including the three aforementioned structures, with a mean absolute deviation of \SI{1.1}{\kcal\per\mole}.
In contrast, all the DFT+MBD approaches (PBE+MBD, PBE0+MBD, and HSE06+MBD) are much closer to FN-DMC, especially for the C$_{60}$@[6]CPPA complex.
The differences between LNO-CCSD(T) and DFT+MBD, and between FN-DMC and DFT+D for the C2C2PD, C3GC and C$_{60}$@[6]CPPA complexes makes it difficult to evaluate a reliable reference interaction energy for these three complexes.
This disparity should motivate the continued development of dispersion-correction methods in order to better characterize the interactions within dispersion-dominated complexes. 

\subsection{MBD properties}

\begin{figure}[t]
\centering
\begin{tikzpicture}
\node[below right, inner sep=0pt] at (0.0,0) {\includegraphics[width=3.25in]{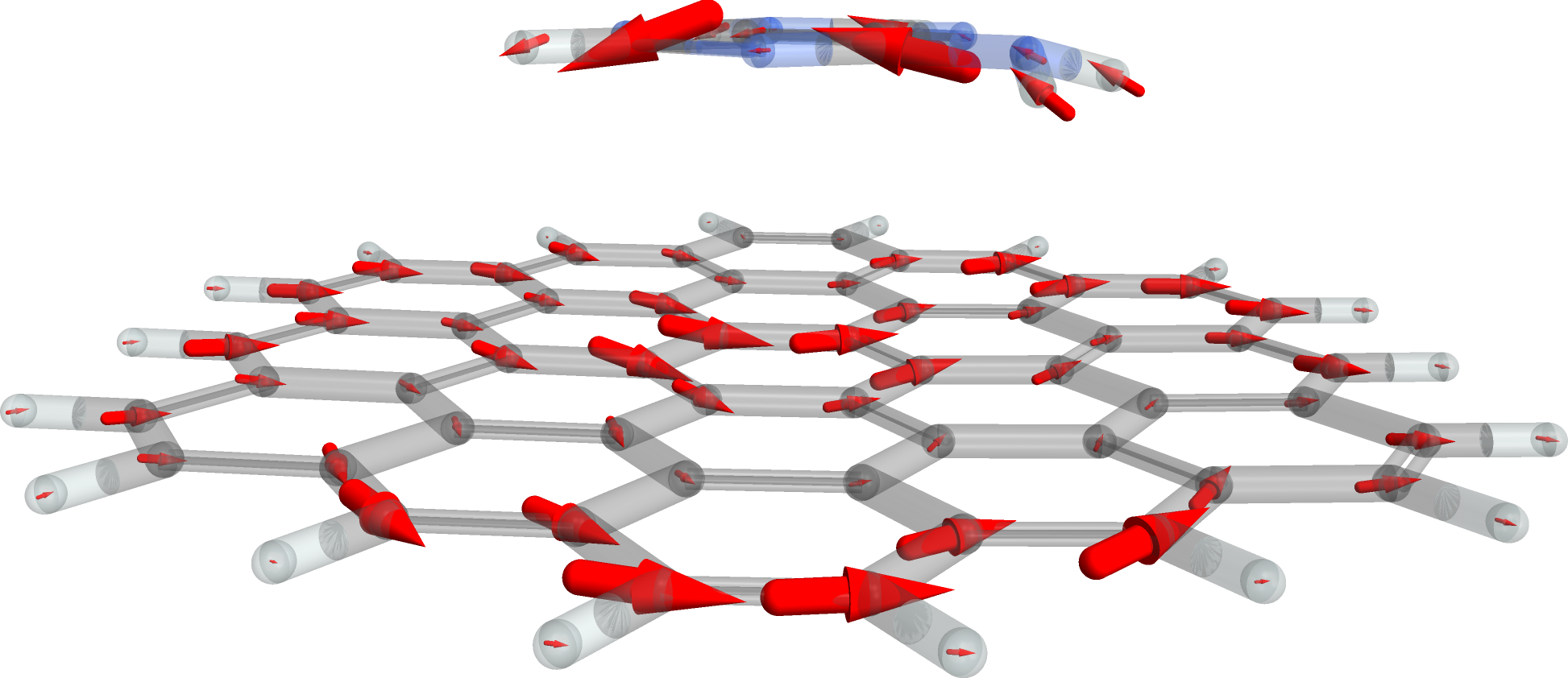}};
\node[below right] at (0,0) {\bfseries a};
\node[below right, inner sep=0pt] at (0.0,-4.5) {\includegraphics[width=3.25in]{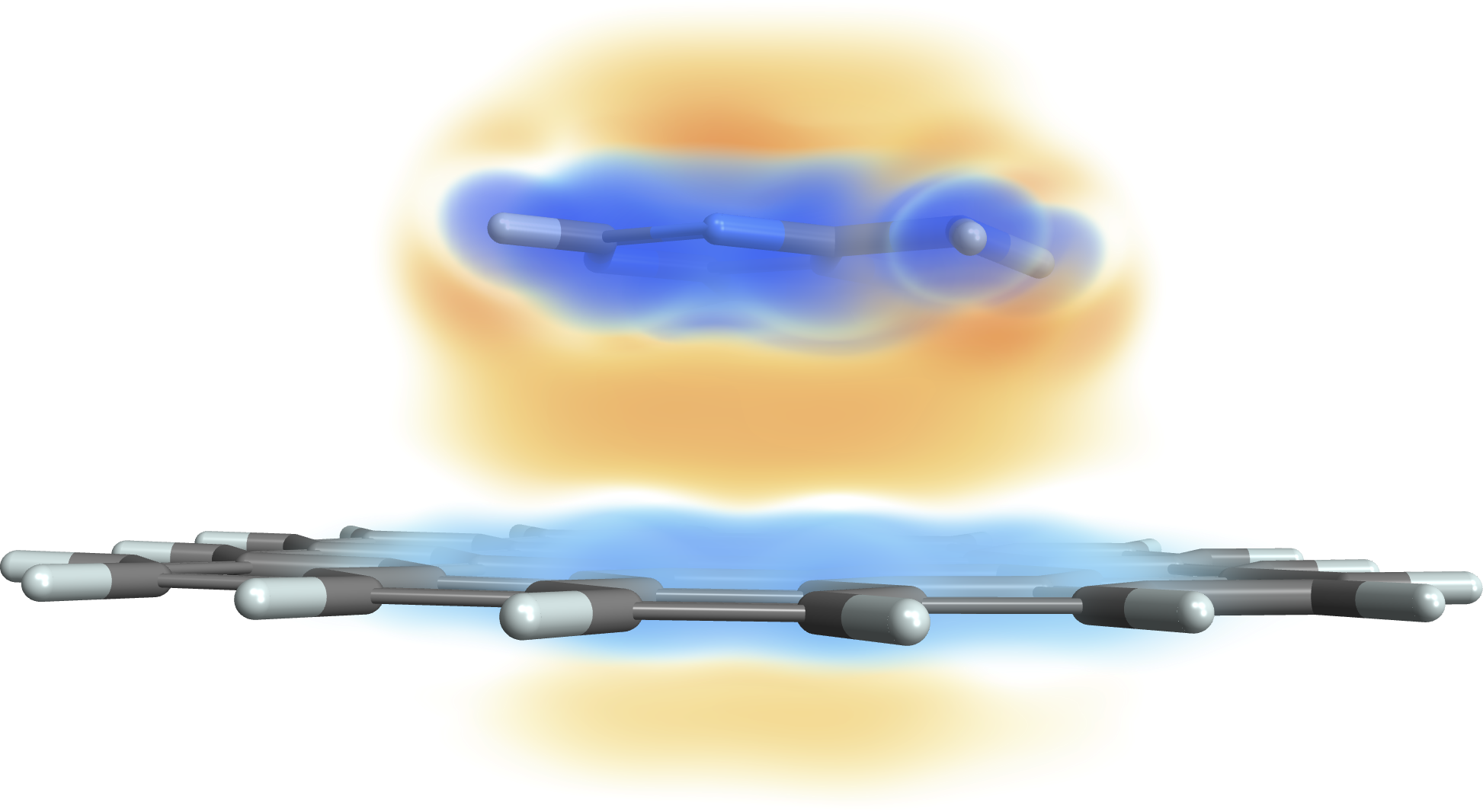}};
\node[below right] at (0,-4.5) {\bfseries b};
\end{tikzpicture}
\caption{\textbf{MBD wavefunction.}
libMBD gives access not only to energies, but also to other expectation values of the MBD wavefunction.
Here calculations on the $\uppi$--$\uppi$-stacked circumcoronene$\cdots$adenine dimer (C3A) of the L7 dataset.
(\textbf a) One of the low-energy zero-point oscillations of the MBD Hamiltonian (\pcref{eq:diagonalization}), displayed as instantaneous dipole displacements of the electronic clouds around each atom.
(\textbf b) Polarization of the electron density due to the vdW interaction (\pcref{eq:density}), with density accumulation and depletion colored as orange and blue, respectively.
}\label{fig:mbd-wf}
\end{figure}

The MBD method is based on an effective Hamiltonian and as such provides access to more than just energies (\cref{sec:mbd-properties}).
The coupled zero-point oscillations of the electron density and the density polarization have been used to reinterpret $\pi$--$\pi$ stacking \citep{HermannNC17,Carter-FenkPCCP20}.
Both can be readily calculated with libMBD using the pyMBD interface.

The C3A dimer from the L7 dataset exhibits typical behavior of these quantities in a $\pi$--$\pi$ complex.
The low-energy zero-point MBD oscillations display long-range order spanning the whole complex, and corresponding to simple electrostatic patterns such as a dipole moment across a whole molecule interacting with an opposite dipole moment on the other molecule (\cref{fig:mbd-wf}a).
The vdW interaction in general polarizes the electron density out-of-plane in $\pi$--$\pi$ complexes.
The C3A dimer demonstrates that while the coupled oscillations are completely delocalized even in the case of differently sized interaction partners, the density polarization is mostly localized in the overlap region between the two molecules (\cref{fig:mbd-wf}b).

\subsection{Previous applications of MBD}

The MBD scheme has been used to identify chemical and thermodynamic trends across a broad range of materials and complexes, and has often succeeded even when other vdW-inclusive methods have failed \citep{StohrCSR19, XuCR20}.

On many occasions, MBD has been able to accurately predict the thermodynamic stabilities of polymorphs often exhibited by molecular crystals, which are typically governed by non-covalent interactions \citep{StohrCSR19, XuCR20}.
The ability to correctly identify the energetic rankings of such vdW-bound systems has applications in numerous fields such as pharmacy and organic electronics \citep{StohrCSR19, BucarACIE15}.
One well-studied case is oxalic acid \citep{StohrCSR19}, for which many vdW-inclusive methods do not correctly predict the relative stabilities of its two polymorphs, whereas PBE+MBD agrees with experimental results \citep{ReillyCS15}.
Another example is coumarin, where the inclusion of beyond-pairwise interactions significantly improves the predicted energy ranking of crystalline polymorphs compared to the pairwise TS method \citep{ShtukenbergCS17}.
Furthermore, the explicit accounting for many-body interactions elucidates the relative prevalence of a particular aspirin polymorph (`form 1') over the other in nature, whereas most other electronic structure methods (both with and without vdW corrections) predict both polymorphs to be energetically degenerate \citep{ReillyPRL14}.

MBD has also been shown to accurately calculate interaction energies for supramolecular host--guest complexes \citep{StohrCSR19}.
In particular, DMC calculations show that the binding energies of the C\textsubscript{70}-fullerene to [10]- and [11]-cycloparaphenylene are degenerate (within DMC uncertainty) \citep{HermannNC17}.
Only the explicit inclusion of many-body interactions alongside DFT correctly predicts this degeneracy, whereas pairwise or two- and three-body vdW models show a clear preference for the 10-membered ring \citep{HermannNC17, AntonyCC15}.


Other applications that the MBD approach has been used for include layered materials \citep{BuckoJPCM16} and molecular switches \citep{ChenJPCC19}.
In the former case, the heat of adsorption of toluene on graphene, as calculated using PBE+MBD, was found to be in good agreement with experiment \citep{BuckoJPCM16}.
Furthermore, the balance between Pauli repulsion, chemical binding, electrostatic interactions, and MBD forces was observed to be critical for achieving molecular switches between the two bistable configurations of gold hexamers on single-walled carbon nanotubes, while the many-body effects were controlled by the former two interactions \citep{ChenJPCC19}.

Many-body dispersion effects have also been shown to be important for the description of  hybrid organic--inorganic interfaces where MBD clearly outperforms additive pairwise dispersion schemes.
For example, MBD was shown to improve the adsorption energetics and geometries of adsorbates such as xenon, graphene, 3,4,9,10-perylene-tetracarboxylic dianhydride (PTCDA), and 7,7,8,8-tetracyanoquinodimethane (TCNQ) on silver surfaces, as they reduce the overbinding typically found in pairwise additive dispersion-correction schemes \citep{MaurerJCP15, BloweyACSNano20}.
Recently, PBE+MBD-NL provided an accurate prediction of the energetic ranking and the adsorption height of different adsorption phases of TCNQ on Ag(100) when compared to X-ray standing wave measurements \citep{SohailJPCC23}.  MBD also permits changes in the anisotropic polarizability tensor, the description of adsorbate vibrations and adsorbate-surface interaction screening to be captured \citep{MaurerJCP15}.
Furthermore, MBD interactions were shown to be a key factor behind the stabilization of a PTCDA molecule being brought to a standing configuration on Ag(111) surfaces using a scanning probe microscopy tip \citep{KnolSA21}.
At a temperature of \SI{5}{\kelvin}, the MBD-calculated lifetime of the standing configuration was calculated to be \SI{1.7E5}{\second}, which was consistent with experimental data; however, the vdW\textsuperscript{surf}-calculated lifetime was evaluated to be only \SI{0.1}{\second} \citep{KnolSA21}.
In addition, the MBD-calculated energy barrier heights were around 170\% larger than the vdW\textsuperscript{surf}-calculated barrier heights \citep{KnolSA21}.
This is because the screening of long-range vdW interactions via non-additive many-body interactions reduces the molecule-surface attraction, which ultimately stabilizes the molecule in the upright geometry due to higher-order contributions to the vdW energy contained in MBD that counteract the pairwise contributions within vdW\textsuperscript{surf}.

\section{Conclusions}
\label{sec:conclusions}

libMBD is a mature, efficient, yet flexible library that offers all the functionality one might need to model vdW interactions in molecules and materials with MBD\@.
It can be used to enhance any electronic structure code that can interface with Fortran, C, or Python external libraries with a powerful implementation of the MBD and TS methods, provided the said code can calculate the Hirshfeld volumes or evaluate the VV polarizability functional (for MBD-NL).
With the recent advent of machine learning models for Hirshfeld volumes and atomic polarizabilities in general, it is even possible to calculate MBD energies with libMBD without running a single electronic structure calculation.
libMBD is hosted and developed on GitHub under the Mozilla Public License 2.0, is distributed with the \citet{ESL-Bundle} \citep{OliveiraJCP20}, and can be easily installed from \citet{conda-forge} and \citet{PyPI}.
We hope that libMBD either standalone or embedded in a growing number of third-party programs will equip researchers with a new computational tool to study molecules and materials.

\begingroup
\setlength\bibitemsep{0pt}
\printbibliography%
\endgroup

\subsection*{Acknowledgments}

\begingroup
\footnotesize
We thank R. DiStasio, A. Ambrosetti, and T. Markovich for countless
discussions about MBD, V. Gobre for writing the original FHI-aims
implementation of MBD, and F. Knoop for correcting mistakes in the
equations in the early version of the manuscript. Funding is
acknowledged from the German Ministry for Education and Research (Berlin
Institute for the Foundations of Learning and Data, BIFOLD), the EPSRC
Centre for Doctoral Training in Diamond Science and Technology
{[}EP/L015315/1{]}, the Research Development Fund of the University of
Warwick, the UKRI Future Leaders Fellowship programme
{[}MR/S016023/1{]}, the European Research Council {[}ERC-CoG grant
BeStMo{]}, and the Luxembourg National Research Fund {[}grant
BroadApp{]}. We are grateful for computing resources from the Scientific
Computing Research Technology Platform of the University of Warwick, the
EPSRC-funded High-End Computing Materials Chemistry Consortium
{[}EP/R029431/1{]} for access to the ARCHER2 UK National Supercomputing
Service (\url{https://www.archer2.ac.uk}), the EPSRC-funded UKCP
Consortium {[}EP/P022561/1{]}, and the partially EPSRC-funded UK
Materials and Molecular Modelling Hub {[}EP/P020194 \& EP/T022213{]} for
access to Young.

\endgroup

%
%

\end{document}